# Deep Radiomics Detection of Clinically Significant Prostate Cancer on Multicenter MRI: Initial Comparison to PI-RADS Assessment

Gabriel A. Nketiah, PhD [a,b,*]; Mohammed R. Sunoqrot, PhD [a,b]; Elise Sandsmark, MD, PhD [b]; Sverre Langørgen [b]; Kirsten M. Selnæs, PhD [a,b]; Helena Bertilsson, MD, PhD [a,c] Mattijs Elschot, PhD [a,b]; Tone F. Bathen, PhD, Prof [a,b]; On behalf of the PCa-MAP Consortium †.

[a] Department of Circulation and Medical Imaging, Norwegian University of Science and Technology, Trondheim, Norway; [b] Department of Radiology and Nuclear Medicine, St. Olavs Hospital, Trondheim University Hospital, Trondheim, Norway; [c] Department of Urology, St. Olavs Hospital, Trondheim University Hospital, Trondheim, Norway; †Names in appendix.

**Correspondence to:** gabriel.a.nketiah@ntnu.no

**Abstract**

*Objective:* MRI radiomics has shown promise in diagnosing prostate cancer, but its practical utility and effectiveness in larger clinical settings remain uncertain. To develop and evaluate a deep radiomics-based model for clinically significant prostate cancer (csPCa) detection and compare its performance to clinical Prostate Imaging Reporting and Data System (PI-RADS) assessment in a multicenter cohort.

*Materials and Methods:* This retrospective study analyzed biparametric (T2-weighted and diffusion-weighted) prostate MRI sequences from four distinct datasets acquired between 2010 and 2020: PROSTATEx challenge, Prostate158 challenge, PCaMAP trial, and an in-house (NTNU/St. Olavs Hospital) dataset. With expert annotations as ground truth, a deep radiomics model was trained, which included nnU-Net segmentation of the prostate gland, voxel-wise radiomic features extraction, extreme gradient boost classification and post-processing of tumor probability maps into csPCa (grade group ≥ 2) detection maps. Training involved 5-fold cross-validation using the PROSTATEx, Prostate158 and PCaMAP datasets, and then externally tested on the in-house dataset. Patient- and lesion-level performance were compared to PI-RADS assessment using area under receiver-operating characteristic curve (AUROC [95% confidence interval]), sensitivity and specificity analysis.

*Results:* Overall, 615 patients (mean age, 63.1±7 years) were evaluated, 199 from PROSTATEx, 138 from Prostate158, 78 from PCaMAP, and 200 from the in-house dataset. In the test data, the radiologist achieved patient-level AUROC of 0.94 [0.91–0.98] with 94% (75/80) sensitivity and 77% (92/120) specificity at PI-RADS ≥ 3 operating point. The deep radiomics model at tumor probability cut-off ≥ 0.76 achieved 0.91[0.86–0.95] AUROC with 90% (72/80) sensitivity and 73% (87/120) specificity, not significantly different (p=0.068) from PI-RADS. On the lesion-level, PI-RADS cut-off ≥ 3 had 84% (91/108) sensitivity at 0.2 (40/200) false positives per patient, whereas deep radiomics attained 68% (73/108) sensitivity at the same false positive rate.

**Conclusion:** A deep radiomics machine learning model achieved comparable performance to PI-RADS assessment in csPCa detection on the patient-level but not lesion-level.

**Introduction**

Prebiopsy MRI has recently emerged as an integral part of the prostate cancer management pathway, enabling MRI-targeted biopsy sampling with associated benefits [1, 2]. However, this is highly hinged on efficient and accurate evaluation of the acquired images. The Prostate Imaging Reporting and Data System (PI-RADS) [3] was introduced to ensure standardized evaluation and interpretation of prostate MRI data. Despite its success, PI-RADS evaluation remains challenging due to the high level of expertise required, its qualitative nature, and signal abnormalities within benign prostate lesions such as benign prostatic hyperplasia and prostatitis, which can mimic or camouflage prostate cancer [4, 5]. Moreover, it remains to be seen if manual PI-RADS evaluation can be scaled up to the projected increase in imaging demand over coming years due to the increasingly aging population and the perceived benefits and wider availability of MRI as a tool for screening. The use of computer aided diagnosis systems involving AI has been touted as a potential solution to the above challenges. The potential of AI, especially that of deep learning approaches in prostate cancer detection and classification, has been shown in several studies [6–9], some of which have reported comparable performance to the radiologist [6, 9]. However, despite their good performance, deep learning methods lack explainability and are often likened to a "black box" of which the "internal behavior" is not fully understood. This could thwart their clinical adoption due to transparency concerns from both clinicians and patients, as well as regulatory compliance challenges [10, 11]. Hence, the concept of explainable AI as a component of deep learning or AI model development has recently gained traction [12].

Unlike deep learning, radiomics constitutes automatic high-throughput extraction of handcrafted quantitative image features (i.e. radiomic features) from radiological images and their subsequent analysis [13], typically using classical machine learning approaches. Radiomics-based machine learning models are easier to understand and interpret in that radiomic features often represent morphological, biological, or pathophysiological phenomena. The potential of MRI radiomics-based machine learning models in prostate cancer diagnosis has been shown in several studies [5, 14–16]. However, most of these studies are based on relatively small patient cohorts from single institutions, have limited practical clinical utility because they are volume-based analysis and not fully automated, and/or lack comparison with to clinical assessment standards. This multicenter study aimed to develop and assess the clinical utility of a fully automated, combined deep learning and radiomics-based machine learning approach for detecting clinically significant prostate cancer (csPCa) and to compare its performance with clinical assessment based on PI-RADS.

**Materials and Methods**

*Dataset, Study Sample and MRI Protocol*

This retrospective study was performed using biparametric (T2-weighted and diffusion-weighted) prostate MRI sequences from four independent datasets: two publicly available training datasets from the PROSTATEx (N=204) [17] and the Prostate158 (N=139) [18] challenges, a collaborative dataset (N=128) from the Prostate Cancer localization with multiparametric MRI approach (PCaMAP) trial (ClinicalTrials.gov Identifier NCT01138527) [15, 19], and an in-house collected dataset (N=248) from NTNU/St. Olavs Hospital, Trondheim, Norway. The datasets were collected in 2012, between February 2016 – January 2020, June 2010 – August 2015, and March 2015 – December 2017, respectively. It consisted of men with clinical indication for prostate MRI examination with a subsequent biopsy (PROSTATEx, Prostate158 and in-house datasets) or surgery/prostatectomy (PCaMAP dataset).

Image acquisition was performed without an endorectal coil on three different types of Siemens (Siemens Healthineers, Erlangen, Germany) 3T MRI systems: Magnetom Skyra, Trio or VIDA. All acquisitions were performed in accordance with institutional and international (PI-RADS) guidelines. Details of the acquisition parameters are shown in supplementary **Table S1**.

Inclusion criteria for this study were complete image acquisition and available pathology report/findings (i.e., ground truth). Patients with history of treatment for prostate cancer, severe image artifacts were excluded. Also, MRI and biopsy negative patients in the in-house data with < 3 years follow-up were excluded. The Regional Committee for Medical and Health Ethics, Mid-Norway, approved this study and the use of the in-

3*Nketiah et al.: Clinically Significant Prostate Cancer Detection using Deep Radiomics*

house dataset (identifier 2017/576) and waived the requirement for written informed consent. Approval was obtained from the PCaMAP trial consortium review board the use of the PCaMAP dataset.

*PI-RADS Reading and Histopathology Assessment*

PI-RADS assessment and biopsy/prostatectomy specimen preparation and evaluation were made locally at each institution in accordance with the International guidelines [3, 20]. The PROSTATEx, Prostate158 and in-house datasets were read by or under the supervision and in consensus with expert radiologists (with $\geq$ 20, 8 and 10 years' experience respectively), who indicated suspicious findings or lesions based on PI-RADS v2 guidelines. Lesions considered likely to be cancerous (PI-RADS score $\geq$ 3) were referred to biopsy. For the PCaMAP cohort, each patient underwent radical prostatectomy within 12 weeks after the MRI examination.

*Lesion Delineation on MRI*

For each dataset, the lesion volumes were retrospectively segmented based on clinical reports from PI-RADS, biopsy or prostatectomy. Lesion delineations of the Prostate158 dataset were publicly available, which were performed independently by two board-certified radiologists with 6 and 8 years of experience. For the PROSTATEx and PCaMAP datasets, we used the same delineations as previously described in [15, 21]. A resident radiologist ($\geq$ 2 years' experience) at St. Olav's Hospital, Trondheim, under the supervision and in consensus with a senior radiologist ($\geq$ 10 years' experience in prostate MRI) delineated the in-house dataset using ITK-SNAP software (version 3.6.0, 2017). It included all MRI-visible (PI-RADS) and histopathologically (biopsy/radical prostatectomy) confirmed lesions.

Lesions with Gleason grade group $\geq$ 2 [9] were labeled as clinically significant prostate cancer (csPCa).

*Image Preprocessing & Feature Extraction*

A 3D nnU-Net (v.1.7.0) model [22] was trained (Python 3.9.12; PyTorch 1.11.0) on the T2-weighted images of the PROSTATEx dataset and then used to segment the whole prostate gland, the peripheral and transition zones for all cases. All subsequent feature extraction (2D voxel-wise) and analysis were performed (in Python 3.8.16) on the automatically delineated whole prostate volumes.

The T2-weighted images were corrected for intensity non-uniformity and non-standardness using N4 bias field correction [23] and dual-reference tissue normalization [21], respectively, and then resampled to 0.5 $\times$ 0.5 mm in-plane resolution. From these, radiomics features (nf = 94, **Table S2**) based on first-, second- and high-order statistics were extracted using PyRadiomics toolkit v.3.0.1 [24].

Apparent diffusion coefficient (ADC) maps and high b-value (b = 1400 or 1500 s/mm$^2$) images were either obtained directly from the diffusion-weighted images series if acquired/generated on-site during scan or calculated off-site from the non-zero b-value (50–800 s/mm$^2$ inclusive) images by monoexponential decay model fitting, as indicated in supplementary **Table S1**. A rigid registration based on Mattes mutual information similarity metric was used co-register and resample (ITK-Elastix toolbox [25]) the ADC maps and high b-value images to match the T2-weighted image resolution. The ADC maps and high b-value images were Gaussian-normalized using global and patient-wise intensity statistics (i.e., mean and standard deviation, respectively), calculated from the segmented prostate volume in the training cohort. First-order statistical radiomics features were then calculated from each normalized image.

Anatomical feature maps consisting of relative distance to the prostate boundary and relative positions in x, y and z directions were calculated from the whole prostate segmentations as defined in [26]. In addition, peripheral zone likelihood maps were obtained directly from nnU-Net.

*Predictive Modelling*

An Xtreme gradient boosting (XGBoost, v.1.7.3) classifier [27], utilizing tree-based booster and logistic regression, was trained to predict the voxel-wise likelihood of clinically significant cancer, generating tumor probability maps (TPmaps). To maintain the multicenter diversity of the data, the classifier was trained on the PROSTATEx, Prostate158 and PCaMAP datasets (n = 415), and subsequently tested on the in-house dataset (n = 200).

For each patient in the training cohort, we selected all ground truth voxels within csPCa lesions, if present, and randomly sampled without replacement non-csPCa (healthy and/or grade group 1 lesions) voxels. Model hyperparameters optimization during training was done using the Optuna (v.3.1.0) optimization framework [28] by minimizing the log-loss across stratified (k = 5)-fold cross-validation on patient-level.

*Post-processing of Tumor Probability Maps*

The TPmaps were post-processed into csPCa detection maps, as defined in the Prostate Imaging Cancer AI (PI-CAI) challenge [29]. Briefly, process this involved thresholding the TPmap voxels based on the Youden's index [30], followed by 3D connected component detection to extract the three largest objects as candidate lesions. For each candidate lesion, the local peak intensity [31] was then calculated using 10 mm diameter spherical kernel, representing the cancer probability of that lesion. csPCa probability threshold of 0.76. which resulted in the highest average of area under receiver-operating characteristic curve (AUROC) and average precision on the training set, was then applied to the test set.

*Model Evaluation*

The performance of the model was evaluated using the standardized functions and metrics for evaluating 3D detection and diagnosis performance in medical imaging as proposed and provided by PI-CAI [29]. Here, a predicted csPCa lesion detection is considered as a hit or true positive if it shares a minimum overlap of 0.10 intersection over union in 3D with the ground-truth annotation. We utilized AUROC for patient-level evaluation, and average precision and free-response receiver operating characteristic curve (FROC) sensitivity for lesion-level evaluation. SHapley Additive exPlanations (SHAP) were used to create feature importance plots, indicating the magnitude and direction of a feature's impact on the model decision for explainability. We evaluated radiologist performance at the PI-RADS operating points and compared the radiologist's AUROC to that of the deep radiomics model using DeLong test [32].

**Results**

*Demographics and Clinical Characteristics of the Study Cohort*

Image data of 719 men from four independent datasets across seven international institutional centers were collected for this study. Out of these, 615 men (mean age ± standard deviation: 64 ± 7 years) met the inclusion criteria (**Figure 1**). This cohort was partitioned into training (PROSTATEx, PCaMAP and Prostate158 datasets; n = 415 [67%]; mean age = 63 ± 7 years) and test (in-house dataset; n = 200 [33%]; mean age = 64 ± 7 years) subsets of independent institutions.

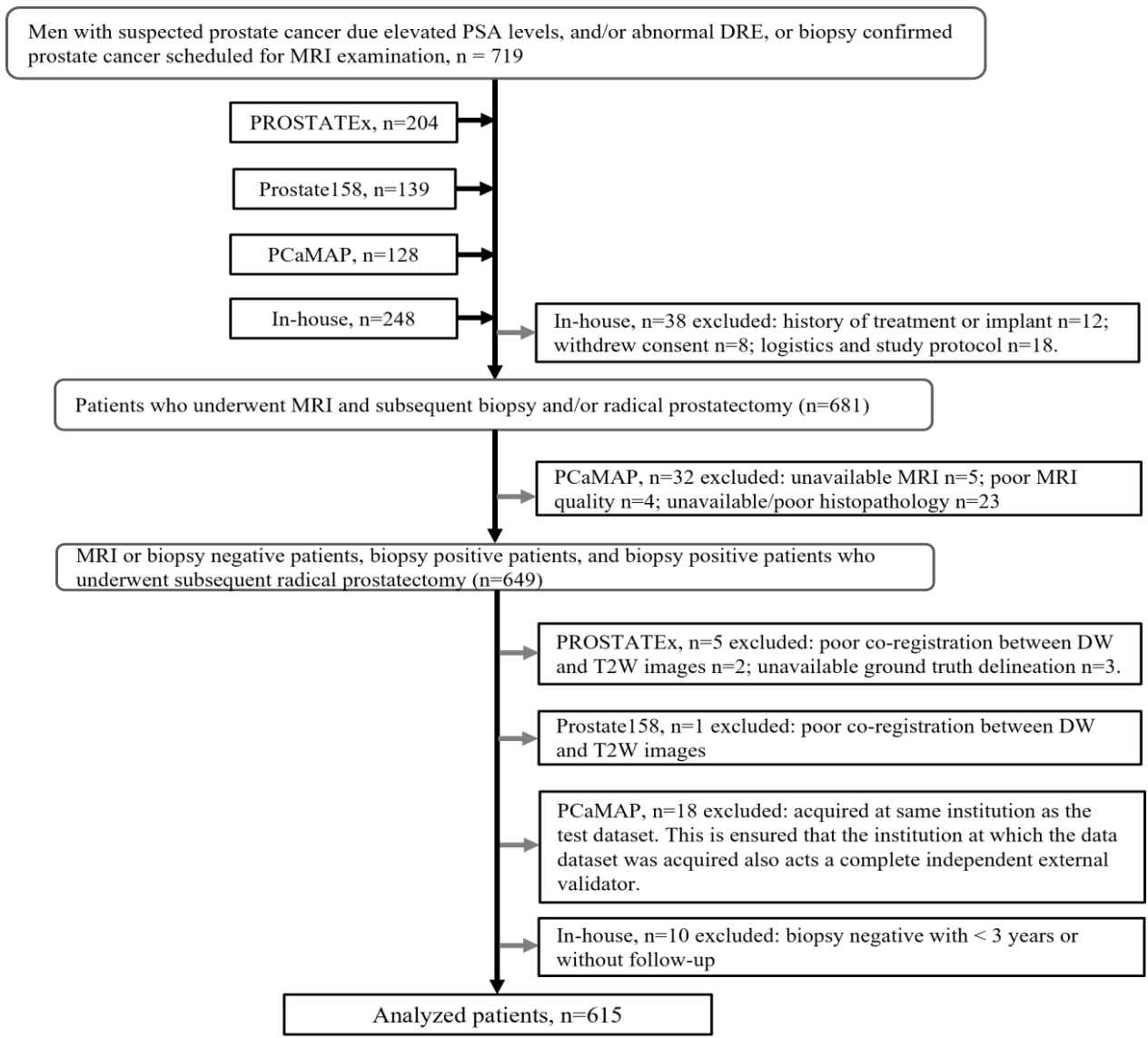

**Figure 1:** Flow chart of patient inclusion in study. PSA = prostate specific antigen, DRE = digital rectal examination, PCaMAP = Prostate cancer localization with multiparametric MRI approach, T2W = T2-weighted, DW = diffusion-weighted.

**Table 1** details the demographics and clinical characteristics of the included cohort. In training cohort, 262 (63%) of the patients had 316 histopathology-confirmed lesions of which 258 (82%) originated from the peripheral zone and 58 from transition zone, and 231 (73%) were csPCa. In the test cohort, PI-RADS assessment reported 131 findings in 103 (52%) patients, consisting of 29 (22%) PI-RADS 3, 34 (26%) PI-RADS 4 and 68 (52%) PI-RADS 5 findings. Ninety-one of these findings in 75 patients were histopathologically confirmed to be csPCa. In total, 131 histopathologically confirmed lesions were found in 93 (47%) patients. One-hundred and eight (82%) of these lesions from 80 patients were csPCa. Among these, 93 originated from the peripheral zone and 15 from the transition zone.

| Table 1: Demographics and Clinical Characteristics of cohort (n = 615) included in study | | | | |
|---|---|---|---|---|
| | Training cohort (n = 415) | | | Test cohort (n = 200) |
| Variable | PROSTATEx (n = 199) | Prostate158 (n = 138) | PCaMAP (n = 78) | In-house (n=200) |
| Mean Age (y) | 63±7 | 66±9 | 61± 6 | 64±7 |
| mean PSA (ng/mL) | 14±10 | 7.5±4.7 | 8±5 | 13.5±21.6 |
| Patients without PI-RADS detected lesions | 62 | 56 | - | 97 |
| Patients with PI-RADS (≥ 3) detected lesions (lesions) | 137 | 83 | - | 103 |
| PI-RADS 3 | - | - | - | 19 (29) |
| PI-RADS 4 | - | 45 | - | 23 (34) |
| PI-RADS 5 | - | 38 | - | 61 (68) |
| Patients with csPCa in PI-RADS detected lesions (lesions) | - | - | - | 75 (91) |
| Patients without pathology-confirmed lesions | 98 | 55 | 0 | 107 |
| Patients with pathology-confirmed lesions | 101 | 83 | 78 | 93 |
| Grade Group per patient (lesions) | | | | |
| 1 | 28 (37) | 9 | 21 (39) | 13 (23) |
| 2 | 38 (41) | 29 | 24 (30) | 25 (40) |
| 3 | 22 (19) | 19 | 19 (20) | 29 (39) |
| 4 | 7 (8) | 18 | 7 (7) | 12 (14) |
| 5 | 6 (7) | 8 | 7 (7) | 14 (15) |
| Pathology-confirmed lesions per patient | | | | |
| One lesion | 85 | 68 | 55 | 62 |
| Two lesions | 12 | 12 | 21 | 27 |
| ≥ Three lesions | 1 | 3 | 2 | 5 |
| Zone distribution of lesions (csPCa) | | | | |
| Peripheral zone | 89 (66) | 79 (79) † | 87 (58) | 113 (93) |
| Transition zone | 23 (9) | 22 (22) † | 16 (6) | 18 (15) |
| Pathology Type | Biopsy | Biopsy or RP | RP | Biopsy = 137; RP = 43 |

PCaMAP = Prostate Cancer localization with a Multiparametric MR Approach, PSA = prostate specific antigen, PI-RADS = Prostate Imaging Reporting and Data System, RP = radical prostatectomy, csPC = clinically significant prostate cancer. Data in parentheses represent lesion-level unless otherwise specified. A lesion originates from a zone of ≥ 75% of the voxels overlap with that zone. † Lesion level grade groups were not available for the Prostate158 dataset.

*Patient-level Performance*

The receiver operating characteristic (ROC) and precision-recall curves in **Figure 2a** and **2c** show the performance of the deep radiomics-based machine learning model in detecting patients with clinically significant cancers compared to the clinical PI-RADS evaluation.

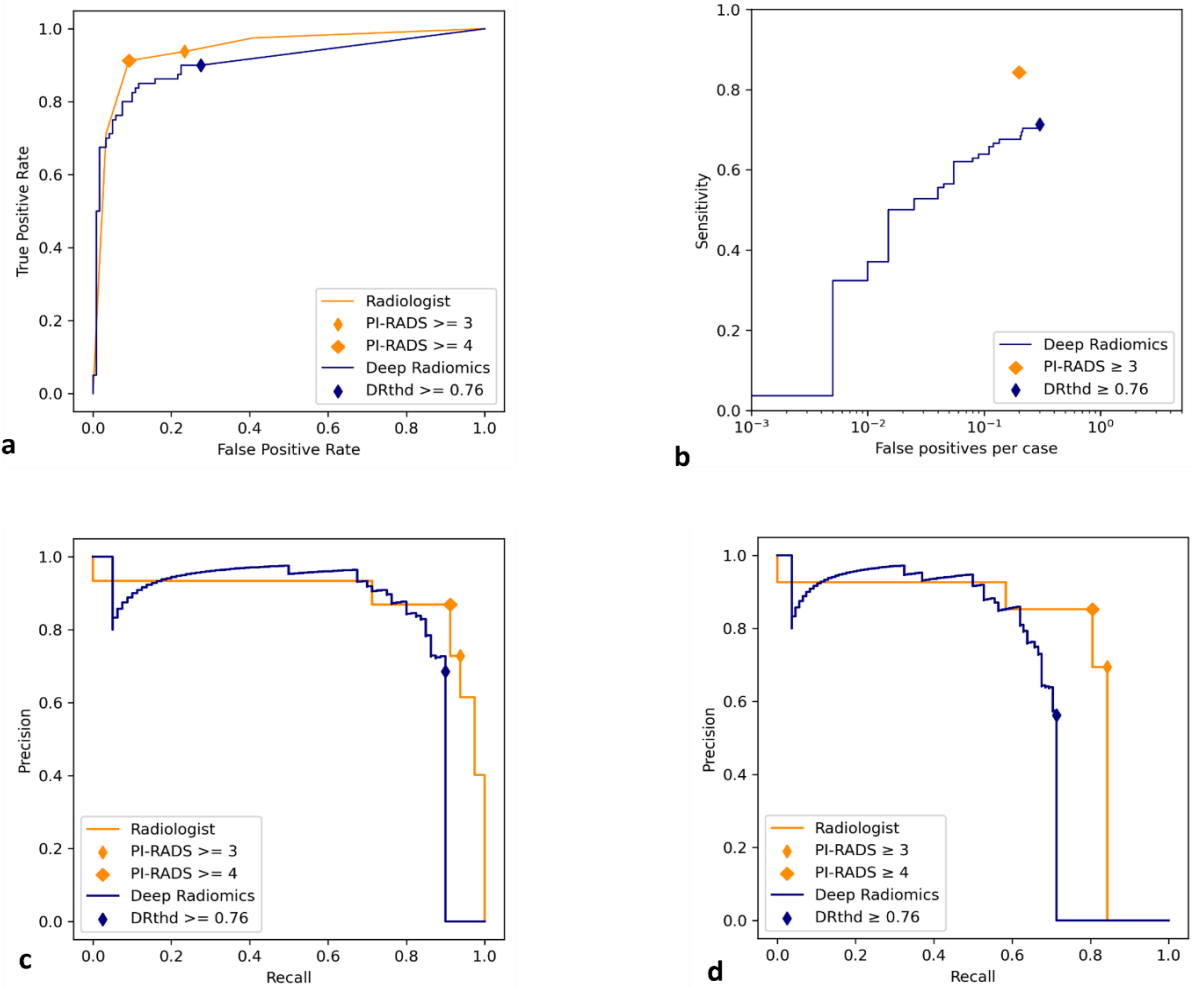

**Figure 2:** (a, b) ROC/FROC and (c, d) precision-recall curves showing the performances of radiologist (orange) and deep radiomics model (navy blue) in detecting clinically significant prostate cancers on (a, c) patient-level and (b, d) lesion-level. PI-RADS = Prostate Imaging Reporting and Data System, DRthd = deep radiomics threshold.

Among the 80 out of 200 patients with histopathology-confirmed csPCa, clinical PI-RADS evaluation achieved a patient-level sensitivity of 94% (75/80) and specificity of 77% (92/120) at PI-RADS threshold of ≥ 3 (**Table 2**). The deep radiomics model achieved 90% (72/80) sensitivity and 73% (87/120) specificity at a tumor probability threshold of ≥ 0.76 (**Table 2**). There was no significant difference (p = 0.0682) in AUROC between PI-RADS evaluation (0.94; CI = 0.91–0.98) and deep radiomics (0.91; CI = 0.86-0.95). At the same specificity as the PI-RADS ≥ 3 operating point (77%), deep radiomics attained a sensitivity of 90% (72/80). The average and thresholded precision scores were 89% and 73% for PI-RADS assessment, and 83% and 67% for the deep radiomics model, respectively.

| Table 2: Diagnostic Performance of Clinical PI-RADS Assessment and Deep Radiomics in Detecting Clinically Significant Prostate Cancers Compared to Histopathologic Mapping Using IoU. | | | | | |
|---|---|---|---|---|---|
| Model, Cutoff | AUROC [95% CI] | Sensitivity (%) | Specificity (%) | FPs Per Case | Precision (%) |
| **Patient-Level** | | | | | |
| Radiologist | 0.94 [0.91-0.98] | | | | 89† |
| PI-RADS ≥ 3 | | 94 (75/80) | 77 (92/120) | 0.23 | 73 (75/103) |
| Deep Radiomics | 0.91 [0.86-0.95] | | | | 83† |
| Thd ≥ 0.76 | | 90 (72/80) | 73 (87/120) | 0.28 | 69 (72/105) |
| Thd ≥ 0.77 (PI-RADS ≥ 3) | | 90 (72/80) | 73 (87/120) | 0.23 | 69 (72/105) |
| **Lesion-Level** | | | | | |
| Radiologist | | | | | 76† |
| PI-RADS ≥ 3 | | 84 (91/108) | | 0.2 (40/200) | 69 (91/131) |
| Deep Radiomics | | | | | 64† |
| Thd ≥ 0.76 | | 71 (77/108) | | 0.3 (60/200) | 56 (77/137) |
| Thd ≥ 0.80 (PI-RADS ≥ 3) | | 68 (73/108) | | 0.2 (40/200) | 65 (73/113) |

PI-RADS = Prostate Imaging Reporting and Data System, IoU = intersection over union, AUROC = area under receiver-operating characteristic curve, FPs = false positives, CI = confidence interval.

32/40 of the false positive PIRADS lesions are from true negative patients (n = 28). 47/60 of the deep radiomics predicted false positive lesions are from true negative patients (n = 33). † Indicates average precision across all thresholds.

*Lesion-level Performance*

**Figure 2b** and **2d** show FROC and precision-recall curves comparing the performances of deep radiomics and clinical PI-RADS assessment in detecting csPCa. The test cohort had 108 pathologically confirmed csPCa lesions present in 80/200 patients. Clinical assessment resulted in 131 candidate lesion detections with 84% (91/108) sensitivity at 0.2 false positives per case for PI-RADS ≥ 3 cutoffs.

Deep radiomics model with optimized tumor probability threshold of ≥ 0.76 resulted in 137 detections with 71% (77/108) sensitivity at 0.3 false positives per case but had 113 detections with 68% (73/108) sensitivity at the same false positives rate as for PI-RADS ≥ 3 (**Table 2**). The average precision scores were 76% for PI-RADS assessment and 64% for the deep radiomics model. Upon thresholding, PI-RADS assessment (≥ 3) and deep radiomics model (tumor probability ≥ 0.76) attained precision scores of 69% (91/131) and 56% (77/137), respectively.

*Feature Importance*

The SHAP plots for a detected csPCa lesion (**Figure 3**) and all detected lesions (**Figure S1**) in the test cohort display the top 20 important features, illustrating their magnitude and direction of impact on the model's decision. First-order statistical radiomic features from ADC and high b-value images comprised majority of the important features. Additionally, significant contributions came from the anatomical features – peripheral zone likelihood (PZL), relative distance to the prostate boundary (RDB), and positional indicators (Xpos, Ypos, Zpos). Only three out of the 94 features from the T2-weighted image were included in this selection.

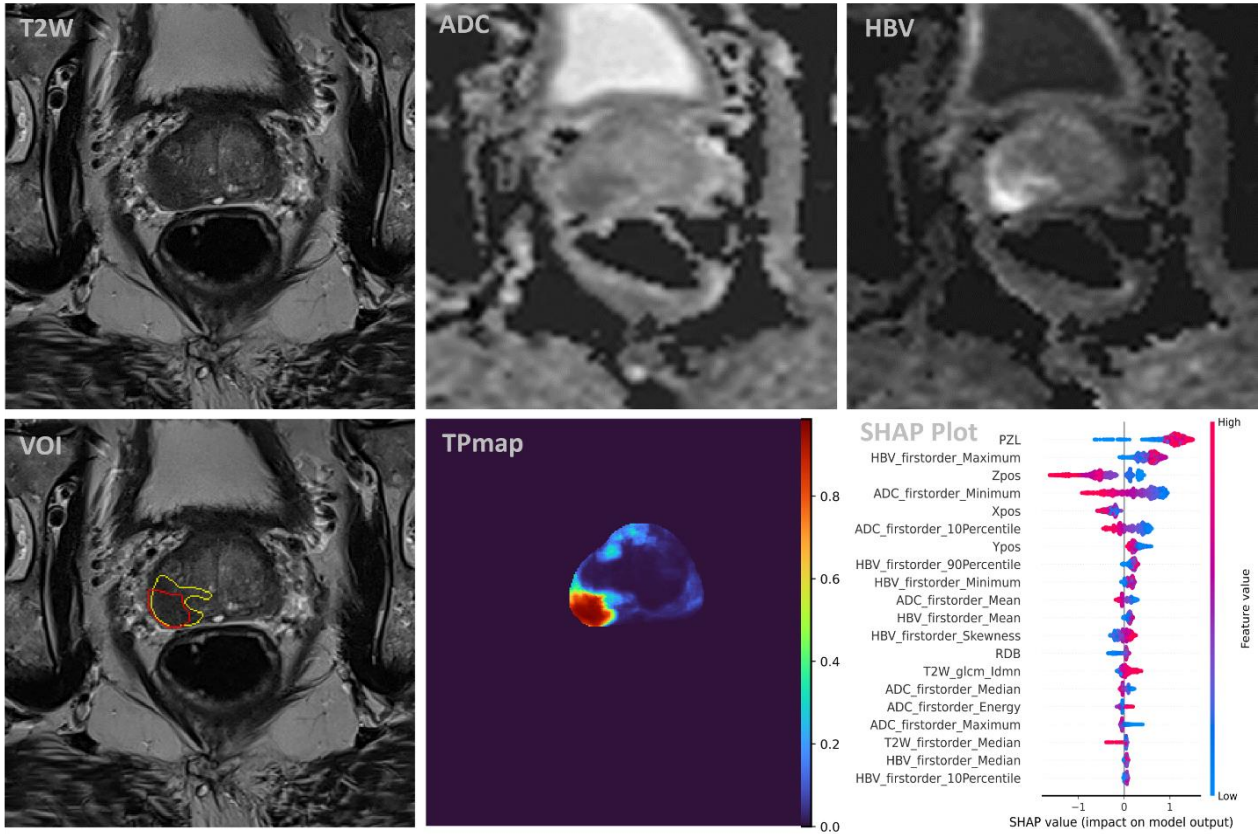

**Figure 3:** Example T2-weigthed (T2W), apparent diffusion coefficient (ADC) map, high b-value (HBV) images and deep radiomics predicted tumor probability map (TPmap) of a 58-year-old test patient with PI-RADS score 5 and grade group 3 (clinically significant) cancer. Ground truth (yellow) and model detected (red) tumor volumes of interest (VOIs) overlayed on the T2W image. The SHapley Additive exPlanations (SHAP) plot from the model detected VOI shows the 20 most important features, along with their magnitude and direction of impact on the model's decision. High or positive SHAP values indicate positive prediction or higher tumor likelihood.

**Discussion**

Diagnostic imaging, especially MRI, plays an important role throughout the entire prostate cancer management pathway. The potential of AI to assist in the analysis of the mounting volume of imaging data and support clinical decision making has been of major interest lately. Several studies have shown the potential of MRI radiomics-based machine learning models in prostate cancer diagnosis [5, 14–16]. However, most of these studies are based on relatively small patient cohorts from single institutions, have limited practical clinical utility and lack comparison with clinical assessment or standards. Whereas deep learning-based methods have been reported [6, 9, 33, 34] to offer high performance, their lack of explainability could hinder their clinical integration. In this relatively large multicentre cohort study, we proposed a practically viable, fully automated and explainable deep radiomics-based machine learning system for prostate cancer detection and compared its performance against routine clinical practice involving PI-RADS assessment. On patient-level, the deep radiomics model achieved a comparable performance (0.91[0.86-0.95] AUROC, 90% (72/80) sensitivity and 73% (87/120) specificity) as PI-RADS assessment (0.94 [0.91-0.98] AUROC, 94% (75/80) sensitivity and 77% (92/120) specificity at cut-off of $\geq 3$), indicating it has potential to assist prostate cancer diagnosis in clinical practice.

Our results are comparable to some of the best performing deep learning methods published [33, 35], but offers more intuition, explainability and transparency throughout the entire pipeline. Hamm et al [35] proposed an explainable deep learning architecture for prostate cancer diagnosis, that is based on training a deep learning (convolutional neural network) model to detect and classify prostatic lesions, followed by training feature classifiers to identify PI-RADS features in the neural activations of the network. While this offers some level of explainability, a deep learning model is directly involved in the actual classification or decision-making, wherein the associations with the PI-RADS features are still learned, and do not represent the importance of the model features directly. In contrast, our approach employs a bottom-up approach, where the radiomic features fed into a classical machine learning classifier are derived from mathematical formulations of morphological, biological, or pathophysiological phenomena, with well-established association to prostate cancer grade group [13–15]. Importantly, SHAP values can be used to directly assess the importance of the radiomics features, along with their magnitude and direction of impact on the model's decision.

Even though deep learning methods may lack explainability, they can still be a valuable component of the radiomics pathway, particularly for tasks that are not directly involved model decision making or where explainability is of less relevance, such as in organ segmentations. This is akin to how deep learning tools are currently applied in clinical practice, where they are used for tasks such as delineating organs for treatment planning or for subsequent expert analysis. In this study, we employed nnU-Net for segmentation of the whole prostate and anatomical feature extraction (e.g., peripheral zone likelihood). This offered two advantages to our model. First, precise delineation of tissue, organ, or volume of interest is an important prerequisite radiomic feature computation and subsequent analysis. By using nnU-Net for segmentation, we removed this mundane manual task and thus enabled complete automation of the analysis pipeline, and also narrowed down the analysis space. Secondly, the zones of the prostate (i.e., transition and peripheral) are known to possess different anatomical, tissue and pathological characteristics, and thus have differences in extracted image features. Most studies therefore recommend zone-wise analysis. However, zone-wise analysis might not be practical or optimal. By incorporating anatomical features such as peripheral zone likelihood, we enabled the model to take into account these distinct characteristics, and thus overcame this challenge. Indeed, these features turned out to be among the top 20 most important features in the model.

The proposed approach is fully automated and could offer several practical advantages to complement radiological workflow with AI assistance. First, it combines multiple MRI series (T2-weighted, ADC, high b-values, etc.,) into a single quantitative feature map (i.e., tumor probability map), which may aid in more efficiency and confidence in detection of csPCa lesions, especially for less experienced readers. Secondly, with comparable patient-level performance as PI-RADS assessment, it could be a suitable tool for initial screening to rule out a subset of patients not requiring biopsy. This could reduce the workload for clinical personnel while still avoiding unnecessary biopsies with associated side effects and cost. Also, if biopsy is required, hotspots within the probability maps can be used to guide sampling and potentially reduce the number of expected biopsy cores. Finally, the extent and direction of feature contribution to the model decision making can be further visualized to explain the model's decisions to the radiologist.

Our study has several limitations. Despite being a multicenter study, the included data is relatively small and comes exclusively from a single-vendor and single-field strength MRI scanner. Compared to other studies [33, 34, 36], our data is less diverse and comes from different clinical populations (i.e. mixture of suspicious, biopsy and prostatectomy patients). This lack of diversity in MRI systems may impact the generalizability of our findings, as MRI signal intensities can vary significantly across different vendor systems and field strengths (e.g., 1.5T vs 3T). This highlights the need for further validation studies using larger and more diverse datasets, including systems from other vendors.

Additionally, the testing was limited to a single independent center (i.e., in-house dataset) because PI-RADS scores and follow-up ($\geq$ 3 years) history were available for only this dataset, which enabled AI comparison with clinical assessment and confirmation of true negative patients. However, the testing lacks a direct comparison with deep learning-based approaches, which limits our ability to contextualize the performance of our deep radiomics method against more advanced AI methodologies. Furthermore, the study is retrospective, and PI-RADS findings were retrospectively matched to histology findings, which may not fully reflect performance in clinical practice.

Finally, a limited number of radiomic features were included in our model despite existence of wide array of texture analysis and radiomic feature extraction approaches [13, 24]. The included features have been extensively studied in different tissues and image modalities and are generally regarded as intuitive. Nevertheless, the merit of including additional feature types is worth investing in future studies.

In conclusion, we presented an explainable deep radiomics machine learning model for detection of csPCa, which achieved comparable performance to PI-RADS assessment on the patient-level but not on the lesion-level.

**Supplemental Material**

| Table S1: Image Acquisition Settings and Parameters | | | | | | | | |
|---|---|---|---|---|---|---|---|---|
| | PROSTATEx | | Prostate158 | | PCaMAP | | In-house | |
| | T2W | DW | T2W | DW | T2W | DW | T2W | DW |
| TR (ms) | 5660 | 2700 | 4040 | 4400 | 4000 | 3300 | 7740 | 4400 |
| TE (ms) | 104 | 63 | 116 | 61 | 101 | 60 | 104 | 63 |
| FOV (mm) | 192 × 192 | 256 × 168 | 180 × 180 | 220 × 220 | 200 × 200 | 260 × 211 | 192 × 192 | 256 × 240 |
| Resolution (mm) | 0.5 × 0.5 | 2.0 × 2.0 | 0.47 × 0.47 | 1.4 × 1.4 | 0.625 × 0.625 | 1.625 × 1.625 | 0.5 × 0.5 | 2.0 × 2.0 |
| Slice thickness (mm) | 3.6 | 3.6 | 3.0 | 3.0 | 3.0 | 3.6 | 3.0 | 3.0 |
| Interslice gab (mm) | - | - | None | None | 0.6 | None | - | - |
| Flip angle (°) | 160 | 90 | 160 | - | 150 | 90 | 160 | 90 |
| B-values (s/mm$^2$) | N/A | 50, 400, 800 | N/A | 0, 100, 500, 1000 | N/A | 0, 100, 400, 800 | N/A | 0, 100, 400, 800 |
| ADC/HBV Calculation | N/A | Off-scanner | N/A | On-scanner | N/A | Off-scanner | N/A | On-scanner |
| Sequence name | TSE | SSEP | - | - | TSE | SSEP | TSE | SSEP |
| MRI systems (3T) | MAGNETOM Trio and Skyra | | MAGNETOM VIDA and Skyra | | MAGNETOM Trio | | MAGNETOM Skyra | |

T2W = T2-weighted, DW = diffusion-weighted, ADC = apparent diffusion coefficient, HBV = high b-value, TR = repetition time, TE = echo time, FOV = field of view, TSE = turbo spin-echo, SSEP = single-shot echo planar.

| Table S2: List of Extracted Radionic Feature Types | | |
|---|---|---|
| **Feature Type** | **Number of Features** | **Source** |
| First-order Statistics | 19 | T2-weighted, ADC, HBV |
| Gray Level Co-occurrence Matrix | 24 | T2-weighted |
| Gray Level Run Length Matrix | 16 | T2-weighted |
| Gray Level Size Zone Matrix | 16 | T2-weighted |
| Neighboring Gray Tone Difference Matrix | 5 | T2-weighted |
| Gray Level Dependence Matrix | 14 | T2-weighted |
| Anatomical | 5 | Prostate mask |

ADC = apparent diffusion coefficient, HBV = high b-value.
Features (non-anatomical) extraction settings: fixed bin width of 10 and a sliding window kernel radius of 2 pixels

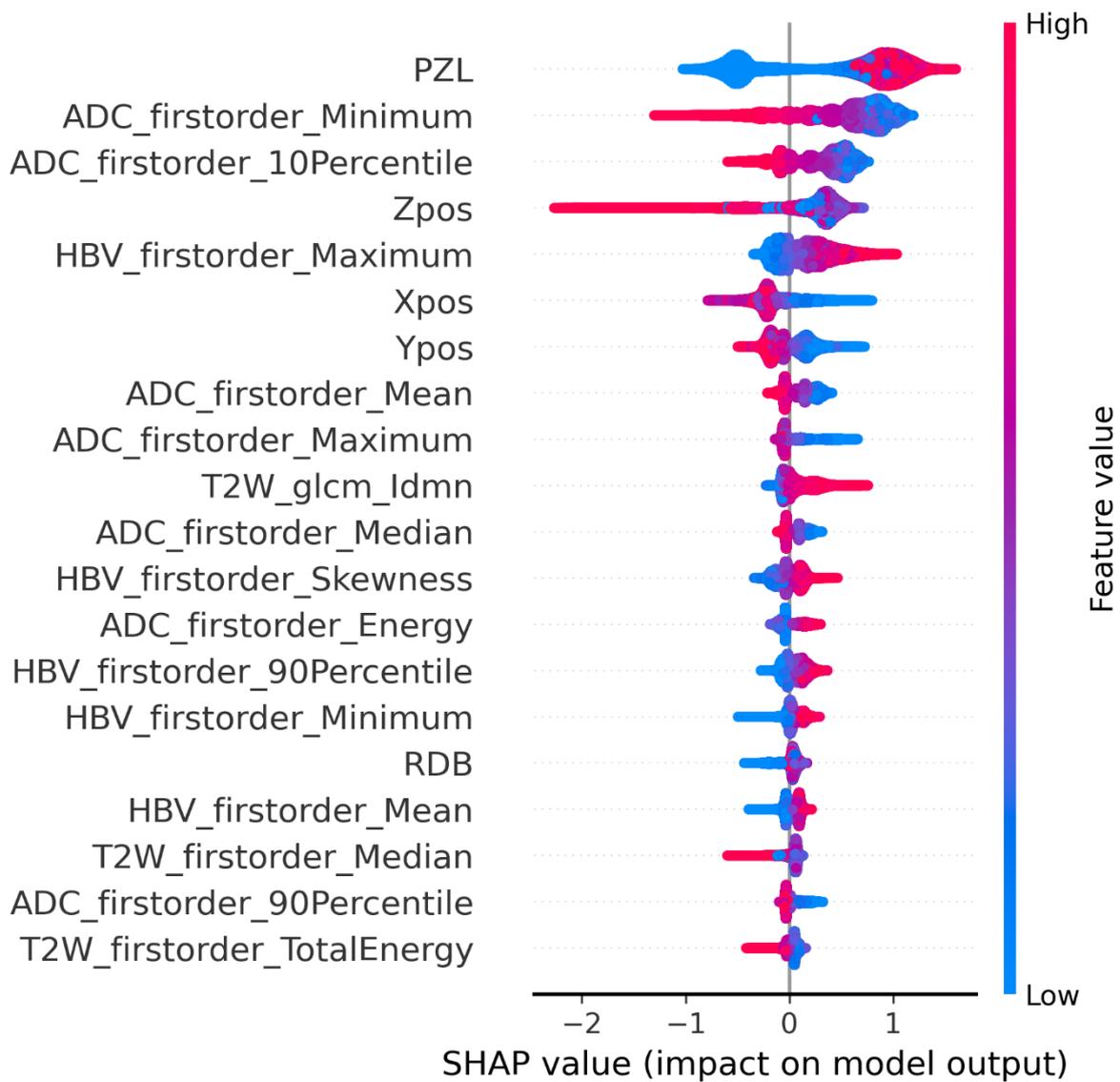

**Figure S4:** SHapley Additive exPlanations (SHAP) plot showing the most important features, along with their magnitude and direction of impact on the model's decision for all lesions detected in the test cohort. High or positive SHAP values indicate positive prediction or higher tumor likelihood.

**The PCaMAP consortium (alphabetical order):**

Ulrike I. Attenberger, MD [a], Pascal A.T. Baltzer, MD [b], Tone F. Bathen, PhD [c], Jurgen J. Fütterer, MD, PhD [d], Masoom A. Haider, MD, FRCP [e], Thomas H. Helbich, MD, MBA [b], Berthold Kiefer, PhD [f], Marnix C. Maas, PhD [d], Katarzyna J. Macura, MD, PhD, FACR [g], Daniel J.A. Margolis, MD [h], Anwar R. Padhani, MD, PhD [i], Stephen H. Polanec, MD [b], Marleen Praet, MD, PhD [j], Tom W. Scheenen, PhD [d], Stefan O. Schoenberg, MD [a], Kirsten M. Selnaes, PhD [c], Theodorus H. van der Kwast [k], Geert M. Villeirs, MD, PhD [j], Trond Viset, MD [l], Heninrich von Busch, PhD [m]

[a] Institute of Clinical Radiology and Nuclear Medicine, University Medical Center Mannheim, Mannheim, Germany; [b] Department of Biomedical Imaging and Image Guided Therapy, Medical University of Vienna, Vienna, Austria; [c] Department of Circulation and Medical Imaging, Faculty of Medicine and Health Sciences; Norwegian University of Science and Technology, Trondheim, Norway; [d] Department of Radiology and Nuclear Medicine, Radboud University Medical Center, Nijmegen, The Netherlands; [e] Dept of Medical Imaging, University of Toronto, Lunenfeld Tanenbaum Research Institute, Sinai Health System, Ontario Institute of Cancer Research, Toronto, Canada; [f] Siemens Healthcare GmbH, MR Application Development, Erlangen, Germany; [g] The Russell H. Morgan Department of Radiology, The Johns Hopkins University, Baltimore, USA; [h] Prostate MRI and Abdominal Imaging Service, Weill Cornell Medicine, Weill Cornell Imaging, New York-Presbyterian, New York, USA; [i] Paul Strickland Scanner Center, Mount Vernon Cancer Center, London, UK; [j] Department of Radiology and Nuclear Medicine, Ghent University Hospital, Gent, Belgium; [k] Laboratory Medicine Program, Princess Margaret Cancer Center, University Health Network, Toronto, Canada; [l] Clinic of Laboratory Medicine, St. Olavs Hospital, Trondheim, Norway; [m] Siemens Healthcare GmbH, AI Products, Forchheim, Germany.